# Hydrogen and Battery – Based Energy Storage System (ESS) for Future DC Microgrids

Massiagbe Diabate, Timothy Vriend, Harish S. Krishnamoorthy, *Senior Member*, *IEEE*

*Abstract*

In this paper, a hydrogen-based energy storage system (ESS) is proposed for DC microgrids, which can potentially be integrated with battery ESS for meeting the needs of the future grids with high renewable penetration. Hydrogen based ESS has the capability to provide a stable supply of energy for a long time, but has slower response compared to battery ESSs. However, a combination of both battery and hydrogen storage provides a stable energy for a long period of time and can easily handle the sudden demands and surplus of the microgrid. One of the main challenges with this system is the integration of power electronics with fuel cells technology to appropriately convert renewable energy into electricity. The proposed system uses an isolated DC-DC converter to activate the production of hydrogen and with the use of electrolyser the produce hydrogen is transformed into hydrogen pressure. The hydrogen pressure becomes an important input to our fuel cell which regulates and transforms it into electricity. The electricity produce is then passed to the different loads by the use of a DC-DC boost converter. In order to verify the effectiveness of the proposed circuit, a 1 kV DC bus voltage hydrogen Simulink simulation is used to demonstrate the hydrogen production and fuel cell behavior based on the demand and surplus power of the loads. The proposed system simulates aspects of the power conversion, electrolyser, storage tank, and fuel cell needed for a complete hydrogen energy storage system. Polymer electrolyte membrane is the main technology focused on for the electrolyser and fuel cell due to its economic feasibility.

## I. INTRODUCTION

As the need for renewable energy grows, energy storages demand increases. One of the problems with renewable energy such as solar and wind is intermittency [1], meaning that these renewable resources are not continuously available to be converted into other energy as they produce power sporadically according to the weather and time of the day. One way to accommodate the power requirements of the grid during times of low power generation is by using energy storage systems [4]. Energy storage technology can help balance the power of the grid, for example during periods of time when the power generated by the renewable energy sources is greater than the power required by the grid, the excess power may be used to charge the energy storage devices such as batteries, capacitors, flywheels, or hydrogen storage. This stored energy can then be used later when the renewable energy sources can not provide the required power. In [5], the author presented a hybrid battery energy storage made of two different types of batteries. 'Energy batteries' are in parallel with 'power batteries' to handle the energy demands and to help reduce the degradation of the energy batteries by sharing the discharging and charging cycle of the power batteries. Different energy storage devices have different strengths and weakness. For instance, capacitors have very high-power density but a relatively low energy density. Lithium batteries have decent energy and power densities, but also rely on lithium metal which is an important resource for other technologies like mobile devices and electric vehicles. Hydrogen storage is one of the most promising storage types due to it high energy density and environmentalism; however, it's power density and efficiency are not very good at this time [3]. In order to take advantage of the unique advantages of the different storage devices, they can be used in conjunction with each other in what is called a Hybrid Energy Storage System (HESS).

This paper proposes a hybrid energy storage system for a 100 % renewable based DC grid/microgrid. This describes the basic control and operation of a hydrogen energy storage system (ESS) in addition to the interaction of the hydrogen storage tank and fuel cell behavior during time of demands and surplus of the loads. The features of the proposed system can be summarized as follows:

- An isolated DC-DC full bridge converter that powers the electrolyzer to generate hydrogen for storage.
- The electrolyzer is controlled to produce the right amount of hydrogen needed to meet the demand and surplus of the loads depending upon the grid requirement.

- Fuel cell transforms the hydrogen pressure into electricity and at the same time, the flow rate is regulated to keep the utilization of the hydrogen just under 100 % to avoid starvation mode and waste as little as possible.
- DC-DC boost converters are used to interface the fuel cell system and battery ESS with the grid.

This paper is organized as follows. In section II, the proposed system will be described, and the hydrogen storage Simulink block will be explained. Section III provides the experimental result and section IV will be the conclusion.

## II. PROPOSED HYBRID ENERGY STORAGE SYSTEM

The overall architecture of the proposed hydrogen- based hybrid ESS is shown in Figure 1. This digest will focus on the operation of the hydrogen ESS; however, the final paper will include the analysis/results from the battery ESS as well. The grid demand is mainly provided by the renewable energy resources (such as solar and wind). Since these are not continuously available and produce electricity intermittently, the hydrogen ESS is implemented to help meet the demands of the load during time of scarce. When, the renewable generation is more than the load demand, the surplus power is transferred to the electrolyzer using a full bridge step-down converter to be stored in the hydrogen storage and during that time the fuel cell is kept inactive. On the other hand, when the renewable generation is less than the grid demand, the stored hydrogen is transformed into electricity by the fuel cell and transferred back to the grid using a DC-DC converter. The proposed system can further be improved by implementing the battery storage, which will result in a hybrid energy storage system. An example of one such combination is to use hydrogen storage along with battery storage such as lithium-ion battery due to their high energy density and low self-discharge. The battery can provide high instantaneous power while the hydrogen cell can produce moderate power for a long duration.

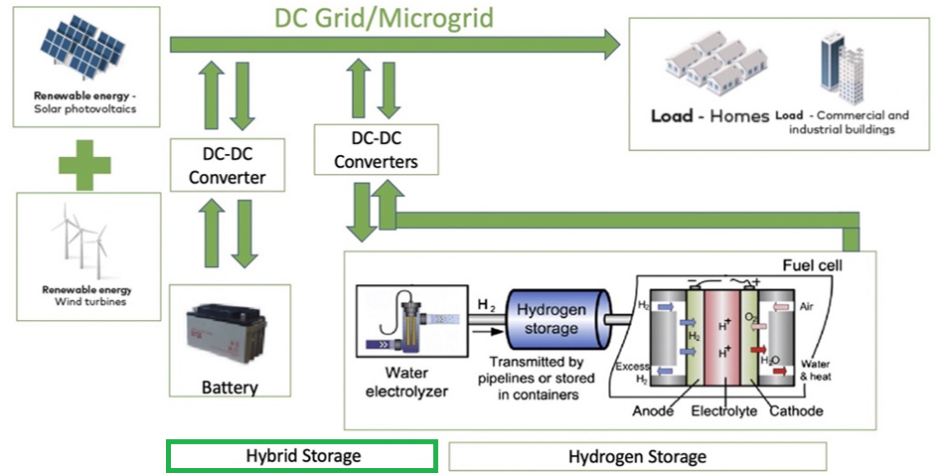

*Figure 1: Proposed Hybrid ESS Architecture for DC Grid/Microgrid*

### A. ELECTROLYZER

An electrolyzer separates water $(2H_2O)$ $into\ hydrogen$ $(2H_2)$ $and\ Oxygen$ $(O_2)$[2]. Hydrogen production in an electrolyzer can be calculated using this following formula: $H_{2\ rate} = \frac{\eta_F * n_{cell} * I_{el}}{m_{H2} * F}$ (1) [1]. Figure 2 shows a simulated model of an electrolyzer. The electrolyzer block is modeled following formula (1) and takes into consideration that the ideal cell voltage is around 1.5 V. To model the ideal cell, two series diodes with forward voltages of 0.75 V each are used. When looking at the cell voltage [V] vs Current Density [A/$cm^2$] diagram, there are ohmic losses as shown in the diagram which changes the cell's voltage with respect to current density. In order to model these losses, a load resistor is added in series with the diodes. The load resistor's value is chosen as the slope of the ohmic region in the I-V curve, which depends on the specific electrolyser's size. Once power is applied to this input impedance approximation, the value of the input current is passed through a conversion formula based on operating temperature and efficiency to arrive at a mass flow rate in moles per second. The step-down converter used to power the electrolyzer uses a perturb-and-observe style power point tracking to capture the excess power produced by the microgrid.

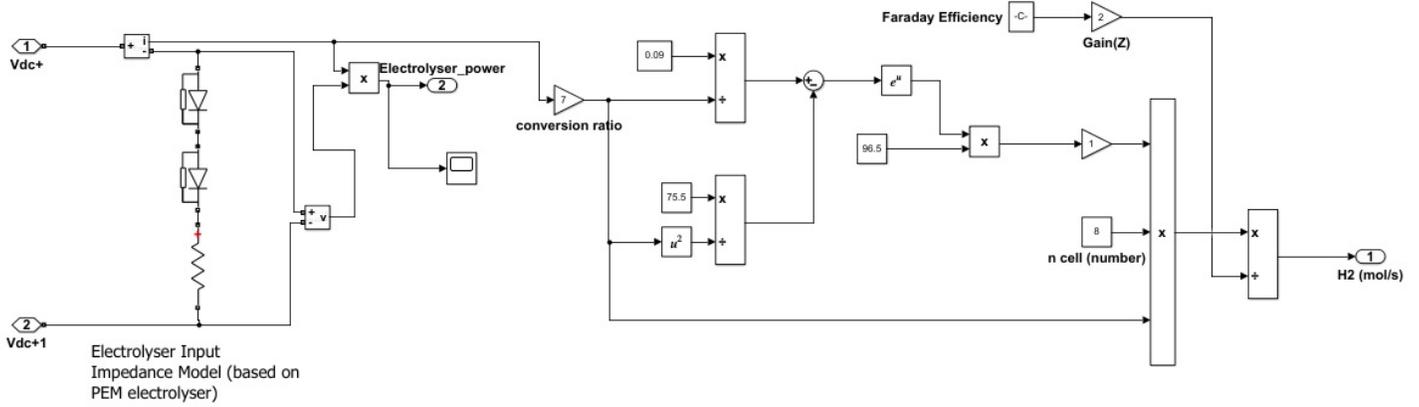

*Figure 2*: **Simulink Model of the Electrolyzer Block**

### B.  HYDROGEN TANK

The hydrogen tank is modeled following the storage tank pressure formula $P_S = P_{si} + \frac{P_{stp}*V_{stp}}{R*T_{stp}} * \frac{\Delta H_{2}\ rate*R*T}{mH_2*V_s}$ (3) [1] in Pascal with the addition of the flow rate fuel cell output from the fuel cell block as a second input in the storage tank as shown in figure 3. The flow rate of the fuel cell is fed back as a second input to the hydrogen tank after converting its unit from $I^3m$ $(Inches^3/m)$ to mol/s using the ideal gas law. Then, the hydrogen produced by the electrolyzer, and the hydrogen taken by the fuel cell are converted into hydrogen pressure.

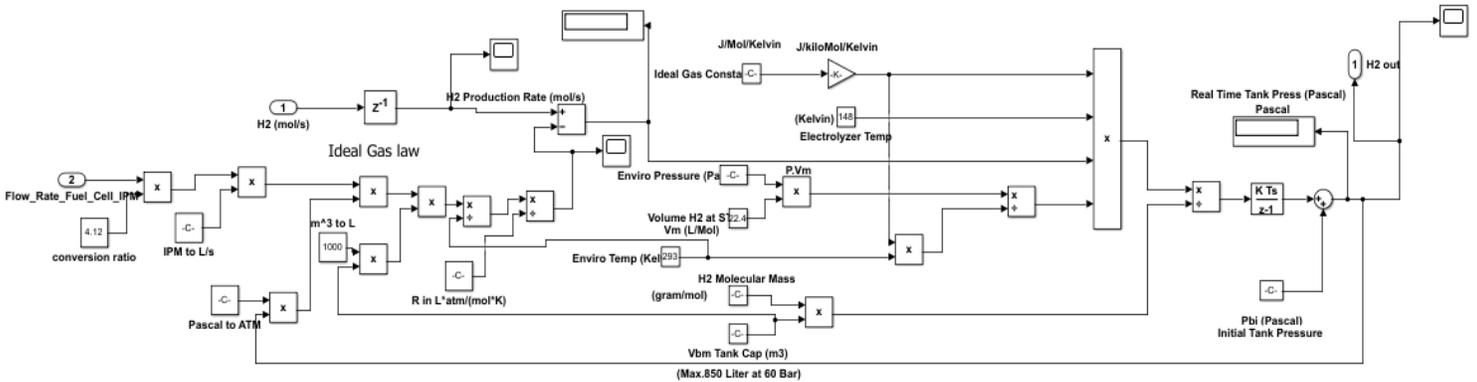

*Figure 3*: **Hydrogen Tank Block**

### C.  FUEL CELL

Figure 4 shows the simulated fuel cell block. For the fuel cell, the built-in 'Fuel Cell Stack' Simulink model is used to model a 6 kW Proton-Exchange-Membrane fuel cell (PEMF). A flow rate regulator feedback loop is added to keep the utilization of the hydrogen just under 100 %. This allows the fuel cell to be kept out of starvation mode and waste as little hydrogen as possible. The flow rate regulator monitors the current produced from the fuel cells and calculates the minimum required amount of fuel to pump into the cell. As long as the fuel cell has sufficient fuel flow rate, the power of the fuel cell is determined by the power drawn by the boost converter. For the control of the boost converter, a perturb-and-observe style controller was used to modulate the converter to match the power demand required by the grid.

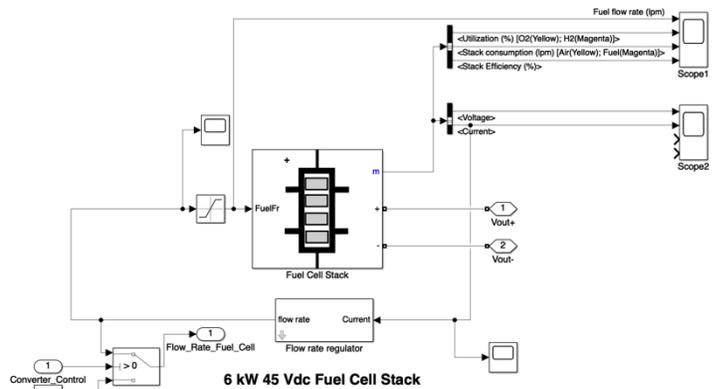

*Figure 4:* **Hydrogen Tank Block**

## III. EXPERIMENTAL RESULTS

A 1 kV DC grid is assumed to analyze the results of the different scenarios explained in section II of the paper. Figure 5 shows the result of the electrolyzer power in green in response to the grid surplus power shown in blue. The electrolyzer power is regulated by the isolated DC-DC converter to be just under the surplus amount because during surplus time, the fuel cell is kept inactive. For a 6 kW surplus, the electrolyzer is regulated at around 5.8 kW and for a surplus of 3 kW, the power is regulated at 2.8 kW that have been implemented. Figure 6 shows the hydrogen production and behavior when all scenarios (demand and surplus) have been implemented. The blue trace is the power surplus (negative value means there is a demand), and the green curve is the hydrogen production. For a surplus of power, the hydrogen stored in the tank increases. When the grid produced exactly the demand of the load the hydrogen tank stops increasing and stays constant. Once the power available to the grid is not capable of meeting the demand of the load, the hydrogen stored in the tank starts decreasing because that hydrogen is now being used by the fuel cell to transform into electricity to meet the demand of the load.

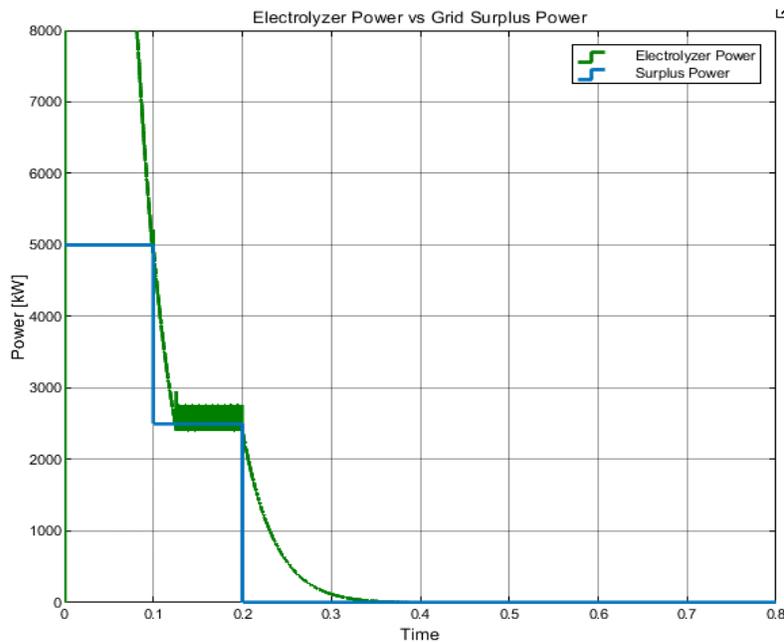

*Figure 5: Electrolyzer Power vs Grid Surplus Power*

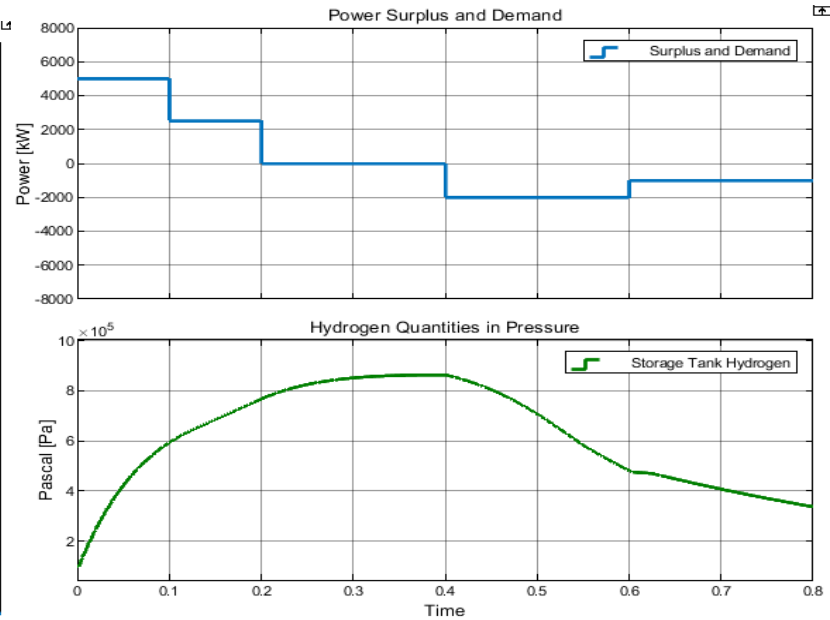

*Figure 6: Hydrogen production vs Surplus and Demand*

## IV. CONCLUSION

This paper proposed a hydrogen-based hybrid energy storage system for a DC microgrid. The paper further discussed the control and operation of a hydrogen ESS in addition to the interaction of the hydrogen storage tank and fuel cell behavior during times of low and high load demand. The Simulink model blocks for the electrolyzer, storage tank, and fuel cell were explained in detail. The proposed control strategy allowed powering of the electrolyzer in case of an excess of energy provided to the grid and activation of the fuel cell when the grid couldn't meet the load demand. The final paper will include the results from battery ESS, including power converter waveforms and the analysis of further benefits of the hybrid ESS during sudden changes in load demand.